\documentclass[conference]{IEEEtran}
\IEEEoverridecommandlockouts
\usepackage{cite}
\usepackage{amsmath,amssymb,amsfonts}
\usepackage{algorithmic}
\usepackage{graphicx}
\usepackage{textcomp}
\usepackage{booktabs}
\usepackage{multirow}
\usepackage{xcolor}
\usepackage{listings}
\usepackage{minted}
\usepackage{url}
\usepackage{tcolorbox}
\tcbuselibrary{skins}

\definecolor{codebg}{HTML}{F7F7F9}
\definecolor{coderule}{HTML}{D0D0D0}
\definecolor{codecomment}{HTML}{6A737D}
\definecolor{codekeyword}{HTML}{005CC5}
\definecolor{codestring}{HTML}{032F62}

\definecolor{dkgreen}{rgb}{0,0.6,0}
\definecolor{gray}{rgb}{0.5,0.5,0.5}
\definecolor{mauve}{rgb}{0.58,0,0.82}

\def\BibTeX{{\rm B\kern-.05em{\sc i\kern-.025em b}\kern-.08em
    T\kern-.1667em\lower.7ex\hbox{E}\kern-.125emX}}
    
\begin{document}

\title{How well LLM-based test generation techniques perform with newer LLM versions?}


\author{\IEEEauthorblockN{Michael Konstantinou}
\IEEEauthorblockA{
\textit{SnT, University of Luxembourg}\\
michael.konstantinou@uni.lu}
\and
\IEEEauthorblockN{Renzo Degiovanni}
\IEEEauthorblockA{\textit{Luxuembourg Institute of}\\
\textit{Science and Technology} \\
renzo.degiovanni@list.lu}
\and
\IEEEauthorblockN{Mike Papadakis}
\IEEEauthorblockA{
\textit{SnT, University of Luxembourg}\\
michail.papadakis@uni.lu}
}

\maketitle

\begin{abstract}
The rapid evolution of Large Language Models (LLMs) has strongly impacted software engineering, leading to a growing number of studies on automated unit test generation. However, the standalone use of LLMs without post-processing has proven insufficient, often producing tests that fail to compile or achieve high coverage. Several techniques have been proposed to address these issues, reporting improvements in test compilation and coverage. While important, LLM-based test generation techniques have been evaluated against relatively weak baselines (for todays' standards), i.e., old LLM versions and relatively weak prompts, which may exacerbate the performance contribution of the approaches. In other words, stronger (newer) LLMs may obviate any advantage these techniques bring. We investigate this issue by replicating four state-of-the-art LLM-based test generation tools, HITS, SymPrompt, TestSpark, and CoverUp that include engineering components aimed at guiding the test generation process through compilation and execution feedback, and evaluate their relative effectiveness and efficiency over a plain LLM test generation method. We integrate current LLM versions in all approaches and run an experiment on 393 classes and 3,657 methods. Our results show that the plain LLM approach can outperform previous state-of-the-art approaches in all test effectiveness metrics we used: line coverage (by 17.72\%), branch coverage (by 19.80\%) and mutation score (by 20.92\%), and it does so at a comparable cost (LLM queries). We also observe that the granularity at which the plain LLM is applied has a significant impact on the cost. We therefore propose targeting first the program classes, where test generation is more efficient, and then the uncovered methods to reduce the number of LLM requests. This strategy achieves comparable (slightly higher) effectiveness while requiring about 20\% fewer LLM requests. 
\end{abstract}


\section{Introduction}

Recent advances in AI, driven by powerful Large Language Models (LLMs), have revolutionized Software Engineering tasks, including code generation, vulnerability detection, program repair~\cite{LLMsurvey}, and more recently, automatic test generation~\cite{foster:mutation, mhetal:fse24-llm, chattester/10.1145/3660783, WangL0J24, WangHCLWW2024}. Despite suboptimal LLM performance (e.g., hallucination), the effectiveness of LLM-based test generation techniques has been significantly improved by combining advanced prompting strategies (e.g., Chain-of-Thought and Self-Consistency) \cite{abs-2201-11903, 0002WSLCNCZ23} and retrieval augmentation~\cite{gao2023retrieval} with traditional software engineering techniques, such as code coverage~\cite{abs-2403-16218} and symbolic execution~\cite{10.1145/3643769}.

All these techniques were effective to mitigate the limitations of LLMs at that time, mainly GPT 3.5. Considering that recent LLMs have improved their reasoning skills, adaptability to various tasks (Mixture-of-Experts) and increased computational efficiency, we raise the question whether the existing LLM-Based test generation techniques remain effective or rather become obsolete, and just by using simple prompting the LLM can produce quality test suites. 

To answer this question, we evaluate four LLM-based test generation techniques for Java programs (HITS~\cite{WangL0J24}, SymPrompt~\cite{10.1145/3643769}, TestSpark~\cite{sapozhnikov2024testspark} and Coverup~\cite{abs-2403-16218}), representative of the state-of-the-art in terms of effectiveness and diversity of the underlying techniques. In addition, we include LLM-Plain, a zero-shot simple prompting baseline. All five approaches are executed on the latest available models, namely GPT-4o-mini, Llama 3.3 70B, and DeepSeek V3. We carefully selected 6 Java projects, by considering the training date of the models, and run the five approaches to generate test cases on the entire project (i.e., on every class under test). 
We compare the performance of the aforementioned approaches on three dimensions. 
\begin{itemize}
    \item We compare the \emph{effectiveness} of the methods in terms of standard test adequacy metrics~\cite{ZhuHM97}, namely line coverage, branch coverage, and the mutation score~\cite{ColesLHPV16}. 
    \item We compare the methods' \emph{efficiency} (cost) in terms of the number of LLM queries required by each approach to generate the test suites (this represents a real-cost component in the case of the commercial models). 
    \item We study how test effectiveness and involved cost is affected by the targeted \emph{granularity level} at the LLM prompt, i.e., when generating tests for specific methods or for entire classes. 
\end{itemize}

Perhaps surprising, our results show that LLMs can produce relatively effective test suites, outperforming the four state-of-the-art tools. Plain-LLM obtains, in total, 49.95\%, 35.33\%, and 33.82\% on line/branch/mutation coverage, while HITS obtains 42.43\%, 29.49\%, and 27.97\%, SymPrompt 40.95\%, 27.19\%, and 28.40\%, TestSpark 23.65\%, 16.48\%, and 13.88\%, and CoverUP 38.85\%, 25.11\%, and 26.26\%, respectively. This indicates that with new, relatively strong LLMs, test generation methods provide a limited or no advantage to the generation process. 

Our results also show that, across the three LLMs used on the experiments, SymPrompt was the least costly tool to run, making 5,602 requests to the LLMs, on average, followed by TestSpark with 7,023 requests, 
Plain-LLM with 11,815, CoverUP 12,741, and finally HITS, the most costly approach, with 16,735 requests. 

While previous approaches were applied at the method level (i.e. the tools generate tests for each method under test), we also evaluate the Plain-LLM prompt at class-level granularity, i.e., the entire class is given as input to the LLM, for which a entire test suite has to be generated. 
In total, Plain-LLM at class-level granularity obtains 33.53\%, 21.05\% and 22.08\% of line/branch/mutation coverage, but needs only 1,562 requests to the LLMs.
Although the generated class-level suites may look less effective than the method-level ones, we observe that these are indeed complementary. 
When both suites are combined, they reach, in total, 53.67\%, 38.74\% and 36.55\% of line/branch/mutation coverage, and improvement of 7.45\%, 9.65\%, and 8.07\%, respectively, but requiring 13,228 LLM requests. 

Although the empirical results are promising and suggest that recent and future releases will keep improving LLMs' effectiveness, there is a clear need to reduce the cost while preserving effectiveness. 
With this spirit in mind, we explore whether a hybrid approach can be a promising solution for this issue and thus, we adapt the Plain-LLM prompting to first perform the class-level generation, and then the method-level generation, targeting only the methods for which some branches remain uncovered. 
We observe that a hybrid granularity prompting can generate suites that reach, in total, 52.30\%, 38.84\% and 36.76\% of line/branch/mutation coverage, with only 10,548 LLM requests. 
Although these hybrid test suites are almost as effective as the combined test suites, we can save 2,680 requests, which constitute almost \textbf{20\% of cost savings}. 

Finally, we observe a key challenge that prevails and needs to be solved by future LLM-based test generation tools.  
On average, near 20\% of the generated tests do not compile, and almost 43\% of the tests do not pass (at least one assertion fails). 
Fixing this great number of non-compiling and non-passing tests can significantly improve their performance, while reducing the costs.

In summary, the key observations of this paper are:
\begin{itemize}
\item Current LLMs have certainly improved their reasoning and generation skills, and even with simple prompts can produce test suites as effective as the ones generated with well engineered tools in the state-of-the-art. 
\item We observe that test suites generated at the class-level granularity can complement the test suites generated at method-level granularity. 
\item A simple hybrid approach that adapts the granularity-level of the generation (targeting first the classes, and then the uncovered methods) can reduce up to 20\% the number of LLM requests, while preserving effectiveness. This is a promising line for future research that we encourage researchers and practitioners to explore. 
\item Finally, one key challenge that future techniques should aim to address regards the high number of non-compiling and non-passing generated tests that, if fixed, can significantly improve the cost-effectiveness of the LLM-based test generation tools. 
\end{itemize}

\section{Background and Related Work}

ChatTester is one of the first studies that used LLMs to generate unit tests \cite{chattester/10.1145/3660783}. ChatTester instructs the LLMs to generate tests for a given class and then iteratively improves them by providing feedback (the compiler error message) every time an error is thrown, i.e., when the test is not compiling. This scheme improves the test suites by generating 34.30\% more compiling tests and 18.7\% more passing tests. This feedback-driven test repair loop was later adopted by many LLM-based unit test generators~\cite{pan2024aster,abs-2503-14000,11042526,abs-2506-02943,SchaferNET24,AlshahwanCFGHHM24,NanG0025,abs-2406-15743}. 
For instance, ChatUniTest~\cite{chatunitest/10.1145/3663529.3663801} includes a test repair mechanism for simple syntactic errors such as imports and missing semicolons, which are taken from the compilation error message. 

TestART~\cite{abs-2408-03095} is another technique that uses template-based and re-prompting to perform test compilation fixing. It relies on five templates to fix common compilation issues. TestSpark~\cite{sapozhnikov2024testspark} aims to enrich the prompts with some context, i.e., by including the class under test together with the relevant parent classes. CoverUp~\cite{abs-2403-16218} introduced the idea of providing coverage feedback, i.e., uncovered code, to LLMs so that they can effectively generate tests for uncovered code. 

SymPrompt~\cite{10.1145/3643769} prompts LLMs with information drawn from traditional symbolic execution. SymPrompt statically analyzes a method to identify all feasible execution paths, and then prompts the LLM to generate a corresponding test case for each path. PANTA~\cite{abs-2503-13580} identifies uncovered code segments and iteratively prompts the LLMs with candidate execution paths as the target. 

HITS~\cite{WangL0J24} is a tool that goes deeper into a method before generating tests through slicing. Using the "Chain of Thought" prompting strategy, it instructs the LLM to decompose a given method into smaller, logically coherent code blocks known as slices. It then prompts the model to generate a corresponding test class for each slice. 

TELPA~\cite{abs-2404-04966} computes a call graph to identify methods' sequence calls that are then used as context in the prompt for test generation. 
There are also some approaches aiming to generate unit tests to \textit{kill} \textit{mutants} \cite{DBLP:conf/icst/StraubingerKL025}~\cite{abs-2506-02954}. Notably, Mutap~\cite{DBLP:journals/infsof/DakhelNMKD24}, ACH~\cite{abs-2501-12862}, PRIMG~\cite{DBLP:journals/corr/abs-2505-05584} are some of the approaches that have been used to generate unit tests to improve the mutation score of generated tests. These approaches differ from the above ones by providing mutation feedback (live mutants) instead of code coverage. 

In general, each of the above approaches focuses on different aspects to make the unit test generation process more effective. 
In this paper, we select HITS~\cite{WangL0J24}, SymPrompt~\cite{10.1145/3643769}, TestSpark~\cite{sapozhnikov2024testspark} and Coverup~\cite{abs-2403-16218} as representative of the aforementioned approaches and evaluate their performance in recent LLMs. 

Here it must be noted that the experimental evaluation we follow in this paper has \textit{significant differences} with that of the previous studies. These are:
\begin{itemize}
    \item \textbf{We evaluate the selected methods on SOTA versions of LLMs}. Most of the existing methods were evaluated on GPT 3.5, which was one of the best LLMs at the time the papers were published, while we are using stronger more recent LLMs belonging to different families, namely, gpt-4o-mini, Llama 3.3. 70B, and DeepSeek V3.
    \item \textbf{We evaluate the selected methods on a carefully selected set of projects (entire projects not selected classes)}. Interestingly, the related studies where evaluated on a self picked set of Java classes, which may have unintentionally introduce a selection bias. To strengthen the related analysis/results, we replicate these methods by running them on entire projects instead of selected classes, which will give us a better understanding of the capabilities of the studied tools and approaches. 
\end{itemize}

\section{Research Questions}
We start by studying the performance of LLM-based test generation techniques on recent LLMs, and thus ask:

\subsection{RQ1: What is the relative effectiveness of LLM-based test generation techniques when using recent LLMs?} 
To answer to this question we include four state-of-the-art LLM-based test generation tools (HITS~\cite{WangL0J24}, SymPrompt~\cite{10.1145/3643769}, TestSpark~\cite{sapozhnikov2024testspark} and Coverup~\cite{abs-2403-16218}), and one LLM-Plain prompting we implemented to use in our experiments. 
We measure the effectiveness of the generated test suites using standard test adequacy metrics~\cite{ZhuHM97}, namely line coverage, branch coverage, and the mutation score~\cite{ColesLHPV16}. 

Since the effectiveness of these techniques comes at a cost, we wonder: 

\subsection{RQ2: What is the relative efficiency (cost) of these LLM-based test generation techniques?}
To answer to this question we measure the number of LLM queries to determine the cost of each workflow's execution (for commercial models, this represents a real-cost component). 

Finally, since LLM-plain test generation can be applied at different granularity levels, in particular, at method-level or at class-level, we ask:

\subsection{RQ3: How the granularity level (Class vs Method) affect the effectiveness and cost of the LLM-based test generation techniques?} 

To answer to this question, we adapt our LLM-Plain strategy to generate test cases for the entire class under test or only a method definition, and thus, we study how this affect on test suites' effectiveness and the number of LLM requests needed. 

\section{Experimental design}

\subsection{Projects used}

\begin{table}
\centering
\caption{Projects used}
\resizebox{\columnwidth}{!}{
\begin{tabular}{ccccccl}
\toprule
\textbf{Project} & \textbf{Abbr} & \textbf{JDK} & \textbf{\# CUT} & \textbf{\# MUT} & \textbf{Previously used} & \textbf{Seen} \\
\midrule
Binance / Binance connector 2.0.0 & BC  & 8  & 46  & 481  & ChatUniTest & Yes \\
bhlangonijr / Chesslib    & CH  & 11 & 43  & 538  & Gitbug Java & Yes \\
AuthMe / ConfigMe       & CM  & 8  & 75  & 541  & Gitbug Java & Yes \\
AWS / Event-ruler             & ER  & 8  & 53  & 656  & HITS        & No  \\
Flmelody / Windward                & W   & 8  & 78  & 400  & HITS        & No  \\
Apple / Batch-processing-gateway& BPG & 17 & 98  & 1041 & HITS        & No  \\
\midrule
\textbf{TOTAL} &       &     & \textbf{393} & \textbf{3657} &  & \\
\bottomrule
\end{tabular}
}
\label{tab:dataset}
\end{table}

For our experiments, we selected six complete open-source Java repositories, representing, to the best of our knowledge, the largest dataset used to date in studies of LLM-based test generation for Java. We selected repositories previously used in LLM-based test generation studies, including those employed for evaluating ChatUniTest~\cite{chatunitest/10.1145/3663529.3663801} and HITS~\cite{WangL0J24}, as well as the GitBug-Java dataset~\cite{SilvaSM24}. To reduce the risk of data contamination, we deliberately chose repositories such that half are not included in the model’s training set. To distinguish between seen and unseen data, we applied the model’s announced cutoff date, classifying all projects created after October 1, 2023, as unseen. Table \ref{tab:dataset} shows the projects used and the repositories that might be included in the LLM's training set.

Unlike prior studies, which typically evaluate only a subset of classes or methods, our evaluation encompasses all classes within each repository. This comprehensive approach avoids introducing bias through selective sampling and ensures that no approach is inadvertently favored.

\subsection{Metrics used}

\textbf{Effectiveness:}
We evaluate the effectiveness of the generated test suites using standard test adequacy metrics~\cite{ZhuHM97}, namely line coverage and branch coverage. Both metrics are measured using JaCoCo, a widely adopted Java code coverage tool that has been extensively used in prior studies on automated test generation. In addition, we computed the mutation score achieved by the generated test suites using PIT~\cite{ColesLHPV16}, a widely used mutation testing framework, with its default configuration.

\textbf{Efficiency:}
We use the number of LLM queries to determine the cost of the execution of each workflow, since the LLM queries represent the most computationally intensive part of the approaches studied~\cite{chattester/10.1145/3660783}. For commercial models, queries also represent a real-cost component. Please note that failed API requests caused by unexpected external errors (e.g., network timeouts) are excluded from the analysis. Hence, the cost reported represents the exact number of requests that each implementation under evaluation requires. 

\textbf{Significance:}
We employ the Mann–Whitney U test~\cite{mann1947test} to assess whether observed differences between approaches are statistically significant. Following standard practice, we consider results to be statistically significant when the p-value is less than 0.05.

\textbf{Compilation and passing rates:} To compute effectiveness metrics, all failing tests must be excluded. We classify failing tests as either non-compiling or non-passing. Although the evaluated LLM-based tools partially filter invalid tests, we apply a uniform post-processing step for consistency:

\begin{itemize}
\item Remove test methods, classes, or helper classes that do not compile.
\item Remove files that fail to compile entirely (e.g., due to import errors). Minor naming mismatches are automatically corrected.
\end{itemize}

Based on such a filtering mechanism, we compute the compilation rate as such:
\begin{equation}
\text{Compilation Rate} =
\frac{N_{\text{generated}} - N_{\text{non-compiling}}}
     {N_{\text{generated}}}
\end{equation}

For the passing rate, we consider tests that compile and pass. We measure the passing rate as such:

\begin{equation}
\text{Passing Rate} =
\frac{N_{\text{generated}} - \left( N_{\text{non-compiling}} + N_{\text{non-passing}} \right)}
     {N_{\text{generated}}}
\end{equation} \\

\subsection{LLM-based unit test generators}

We used recent state-of-the-art LLM-based test generation approaches to evaluate their current effectiveness. The original evaluations of these approaches were conducted using an older such as GPT-3.5-turbo or GPT-4. In our experiments, we ran each approach with its default configuration, but replaced the underlying model with its successor, GPT-4o-mini. Each of the following approaches is built on top of a variation of the LLM-Plain workflow illustrated in Figure \ref{fig:plain_workflow}, and each approach has a unique generation technique. Namely, we employ the following approaches:

\textit{HITS}~\cite{WangL0J24} uses the LLM to decompose methods into code chunks. Then, it applies the LLM-based unit test generation workflow on each of the code chunk.

\textit{SymPrompt}~\cite{10.1145/3643769} attempts to identify execution paths similarly to symbolic execution. Once all execution paths are identified, SymPrompt will use the LLM to generate unit tests for each path. 

\textit{TestSpark}~\cite{sapozhnikov2024testspark} can use either traditional test generation via EvoSuite or LLM-based test generation. When employing LLMs, \textit{TestSpark} prompts the model to generate test cases targeting 100\% coverage for the entire class or method under test. Additionally, if the class under test inherits from a parent class, \textit{TestSpark} includes the corresponding parent class code in the prompt. In our experiments, we exclusively used the LLM-based configuration, since we evaluate LLM-based tools previously assessed with earlier model versions.

\textit{CoverUp}~\cite{abs-2403-16218} extends prior test generation workflows by introducing an additional coverage augmentation phase. After generating an initial test suite, CoverUp measures code coverage, identifies uncovered lines, and subsequently guides the LLM to generate additional tests targeting the remaining uncovered code.

Although further approaches have been tried so far, we did not include every approach into our evaluation either because the approach does not contain a publicly available implementation (e.g. ASTER~\cite{pan2024aster}), or simply because they have not been evaluated with older versions of the LLMs of this evaluation study (e.g. PANTA~\cite{abs-2503-13580}). Additionally, certain approaches, such as \textit{ChatTester}, were excluded despite having been evaluated with earlier versions of our chosen LLMs, as they have been consistently outperformed by one or more of the selected baseline tools.

\subsection{LLM-Plain}

\begin{figure}[t]
    \centering
    \includegraphics[width=\linewidth]{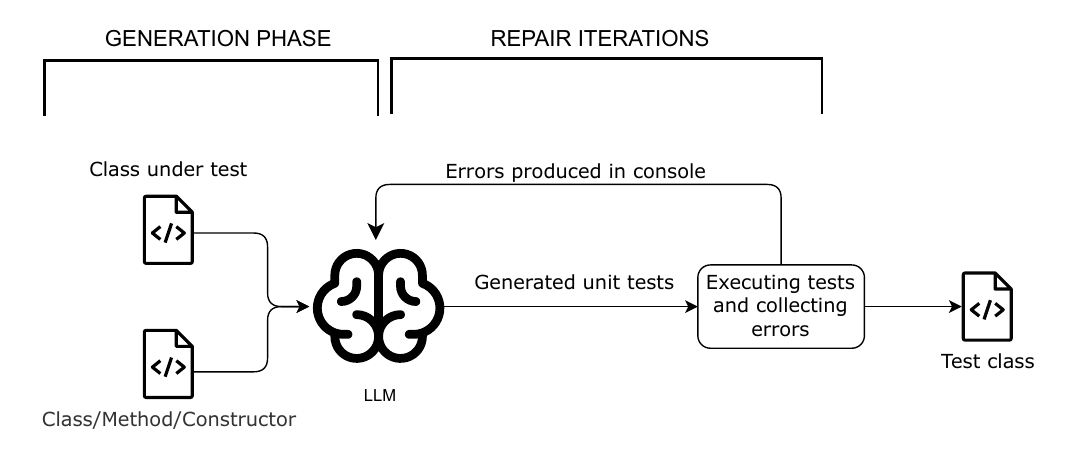}
    \caption{LLM-Plain unit test generation workflow}
    \label{fig:plain_workflow}
\end{figure}

LLM-Plain (not to be confused with "Raw LLM") primarily leverages LLM capabilities with minimal parsing. The aim of this implementation is to provide a simplistic approach in which one could build on top of it.

As shown in Figure \ref{fig:plain_workflow}, the approach uses a single generation prompt to produce unit tests for a given class or method, with the full class code as input. Then, a maximum of five repair iterations is applied, just like in the aforementioned baselines. In each iteration, the generated tests are executed, and any errors are sent back to the LLM for correction. The full communication history is maintained throughout, preserving context across iterations. Minimal parsing ensures that the package name, imports, and test class name match the class under test, guaranteeing syntactic consistency. This minimal parsing is similar to the baselines' rule-based fixing. To reduce nondeterminism, the LLM’s temperature is set to 0.1.

We consider this approach as "LLM-Plain", as it relies mostly on the capabilities of the LLMs and not on any other existing techniques, static analysis or prompt engineering. Depending on the scope on which we run LLM-Plain, the workflow is slightly changed as follows:

\subsubsection{Class-level workflow}

Given the entire class under test, the LLM is asked to generate tests for the whole class into a single test file (e.g. FooTest will be the only generated test file testing class Foo). 
\begin{tcolorbox}[colback=gray!10, colframe=gray!80, boxrule=0.4pt, arc=1pt,
                  left=6pt, right=6pt, top=4pt, bottom=4pt,
                  fontupper=\small\ttfamily, before skip=8pt, after skip=8pt,
                  title=Generation Prompt (Class),
                  enhanced]
The following class is missing unit tests. Please generate all tests needed to achieve 100\% code coverage using Java and Junit5. Return only the code

<class content>
\end{tcolorbox}

\subsubsection{Method-level workflow}

The workflow is similar to the aforementioned class-level testing, but repeated for each method. Given the entire class under test, LLM-Plain executes the generation process for each method name. As a result, LLM-Plain will not generate a test class for each cut, but a test class for each mut. In addition, if the class under test contains at least one constructor, LLM-Plain will repeat the generation process one more time, asking the LLM to generate tests to cover the constructors of the class.

\begin{tcolorbox}[colback=gray!10, colframe=gray!80, boxrule=0.4pt, arc=1pt,
                  left=6pt, right=6pt, top=4pt, bottom=4pt,
                  fontupper=\small\ttfamily, before skip=8pt, after skip=8pt,
                  title=Generation Prompt (Method),
                  enhanced]
The following class is missing unit tests for method <method>. Please generate all tests needed to achieve 100\% code coverage for method <method>, using Java and Junit5.
The name of the generated test class must be <class name>. Return only the code

<class content>
\end{tcolorbox}

\section{Results}

\subsection{RQ1: Effectiveness}

Table \ref{tab:effectiveness} records the code coverage achieved by all the approaches we study. The approach obtained the highest coverage for each project is highlighted in \textbf{bold}, while the second-best performance is \underline{underlined}. At the bottom of the table, we report the overall coverage across the entire dataset, over all included projects. The effectiveness of Plain-LLM at the method level is reported as Plain (M).

\begin{table*}[t!]
\centering
\caption{Effectiveness between HITS, SymPrompt, TestSpark, CoverUp, Plain-Class, and Plain-Method}
\resizebox{\textwidth}{!}{
\begin{tabular}{lcccccc|cccccc|cccccc}
\toprule
\textbf{Project} &
\multicolumn{6}{c}{\textbf{Line Coverage}} &
\multicolumn{6}{c}{\textbf{Branch Coverage}} &
\multicolumn{6}{c}{\textbf{Mutation Score}} \\
\cmidrule(r){2-7}\cmidrule(r){8-13}\cmidrule(r){14-19}
& HITS & SYMP & TSP & CUP & Plain (C) & Plain (M)
& HITS & SYMP & TSP & CUP & Plain (C) & Plain (M)
& HITS & SYMP & TSP & CUP & Plain (C) & Plain (M) \\
\midrule

BPG
& 35.52\% & 27.98\% & 17.05\% & 25.20\% & \underline{36.47\%} & \textbf{45.13\%}
& \textbf{35.89\%} & 30.41\% & 19.04\% & 25.39\% & 22.88\% & \underline{32.60\%}
& 27.35\% & 23.76\% & 14.34\% & 19.62\% & \underline{29.33\%} & \textbf{41.21\%} \\

ER
& 8.96\% & 13.21\% & 15.81\% & 10.66\% & \underline{35.58\%} & \textbf{45.55\%}
& 5.00\% & 8.70\% & 10.13\% & 7.48\% & \underline{23.36\%} & \textbf{32.10\%}
& 6.61\% & 10.56\% & 11.41\% & 9.08\% & \underline{23.03\%} & \textbf{31.53\%} \\

BC
& 71.47\% & \underline{83.02\%} & 18.41\% & \textbf{84.05\%} & 15.21\% & 49.91\%
& \textbf{79.45\%} & 65.75\% & 36.07\% & \underline{70.32\%} & 36.20\% & 49.77\%
& 45.32\% & \underline{49.30\%} & 10.77\% & \textbf{50.12\%} & 7.85\% & 15.93\% \\

W
& \underline{45.28\%} & \textbf{48.82\%} & 17.95\% & 38.70\% & 37.14\% & 36.96\%
& \underline{41.00\%} & \textbf{44.02\%} & 5.98\% & 31.08\% & 24.90\% & 27.49\%
& 32.35\% & \textbf{39.60\%} & 6.84\% & 27.06\% & 30.48\% & \underline{32.81\%} \\

CM
& \underline{62.40\%} & 54.68\% & 34.61\% & 54.22\% & 41.62\% & \textbf{65.39\%}
& \textbf{51.48\% }& 44.07\% & 27.78\% & 44.81\% & 25.56\% & \underline{48.89\%}
& \underline{44.69\%} & 43.01\% & 21.37\% & 43.13\% & 28.50\% & \textbf{53.24\%} \\

CH
& \textbf{65.41\%} & 59.47\% & 44.09\% & 60.56\% & 31.24\% & \underline{61.86\%}
& \textbf{44.31\%} & 38.30\% & 21.88\% & 39.58\% & 8.57\% & \underline{39.10\%}
& \textbf{40.63\%} & 36.20\% & 17.59\% & \underline{36.77\%} & 14.11\% & 30.00\% \\

\midrule
\textbf{Total}
& \underline{42.43\%} & 40.95\% & 23.65\% & 38.85\% & 33.53\% & \textbf{49.95\%}
& \underline{29.49\%} & 27.19\% & 16.48\% & 25.11\% & 21.05\% & \textbf{35.33\%}
& 27.97\% & \underline{28.40\%} & 13.88\% & 26.26\% & 22.08\% & \textbf{33.82\%} \\

\bottomrule
\end{tabular}
}
\label{tab:effectiveness}
\end{table*}

Based on the line coverage results, the Plain-LLM approach consistently outperforms all previously proposed techniques in total and average coverage. Although Plain-LLM does not achieve the highest coverage in every individual repository, its performance remains consistently comparable across projects, resulting in the highest overall effectiveness.

For branch coverage, HITS achieves the highest coverage in the majority of individual repositories; however, when aggregating results across the entire dataset, Plain-LLM covers a larger total number of branches. On average, the differences among Plain-LLM, HITS, and \textit{SymPrompt} are modest. Regarding mutation testing, Plain-LLM again emerges as the most effective approach in terms of total mutation score, with \textit{SymPrompt} ranking second. 

Overall, these results indicate that even without sophisticated error parsing or extensive engineering, a Plain-LLM approach that relies primarily on the inherent capabilities of large language models can outperform prior techniques in terms of line coverage and mutation score, while achieving comparable performance in branch coverage. 

Statistical tests also confirm this observation. Here, it must be noted that since the effectiveness differences are relatively small, the key claim is that the other methods do not offer any advantage over Plain-LLM. 

\begin{tcolorbox}[colback=gray!10, colframe=gray!50,boxrule=0.4pt,arc=1pt,left=6pt,right=6pt,top=4pt,bottom=4pt,fontupper=\small,before skip=8pt,after skip=8pt,enhanced]
\textbf{Finding 1:} LLM-based test generation methods have comparable test effectiveness (in fact, Plain-LLM is slightly superior) with  Plain-LLM.
\end{tcolorbox}

\subsection{RQ2: Efficiency}

Table \ref{tab:requests_per_project} records the number of calls each approach requests to the LLM, that is, the number of API queries required to generate test suites for all repositories in our dataset. Please notet that this number actually represents the most costly part of the approaches. The results show that approaches such as \textit{SymPrompt} and \textit{TestSpark} generally require fewer requests. Notably, the Plain-LLM approach requires fewer LLM queries than both \textit{CoverUp} and \textit{HITS}, while achieving higher test effectiveness.

The higher query cost of \textit{CoverUp} can be attributed to its additional coverage augmentation phase, which triggers additional LLM interactions beyond the initial test generation. In contrast, Plain-LLM does not incorporate such a post-processing step. \textit{HITS}, on the other hand, applies a finer-grained strategy by decomposing methods into smaller code chunks and invoking the test generation workflow for each chunk. Although this enables more targeted test generation, it nearly doubles the number of LLM requests compared to Plain-LLM.

\textit{SymPrompt} is also applied at the method level, but it goes a step further by focusing on generating tests for individual execution paths. This is an attempt to make the LLM focus on different execution for each method by providing specific and concise targets to cover when producing tests. Interestingly, this results in making fewer calls to generate tests than Plain-LLM; however, such a targeted approach does not necessarily give superior test effectiveness.

\begin{tcolorbox}[colback=gray!10, colframe=gray!50,boxrule=0.4pt,arc=1pt,left=6pt,right=6pt,top=4pt,bottom=4pt,fontupper=\small,before skip=8pt,after skip=8pt,enhanced]
\textbf{Finding 2:} Plain-LLM offers a competitive balance between efficiency and effectiveness by making comparable number (fewer) of LLM calls to the other methods. Interestingly,  the second most effective approach, HITS, requires nearly twice as many LLM queries as Plain-LLM does.
\end{tcolorbox}

\subsection{RQ3: Granularity}

Table \ref{tab:class_method_combined} records the coverage achieved by Plain-LLM when instructed to generate tests at the class level in contrast to the method level. To further investigate the observed differences between these two configurations, we additionally merged the passing test suites produced by both class-level and method-level prompting (combined approach). We then measured the combined coverage of both sets of generated tests. \\

\begin{table}[H]
\centering
\caption{Number of Requests to the LLM}
\scriptsize
\resizebox{\columnwidth}{!}{
\begin{tabular}{lcccccc}
\toprule
\textbf{Project} & \textbf{HITS} & \textbf{SymPrompt} & \textbf{TestSpark} & \textbf{CoverUp} & \textbf{Plain (Method)} \\
\midrule
BPG & 3985 & 994  & 967  & 2047 & 2196 \\
ER  & 2530 & 1081 & 1100 & 1579  & 2429 \\
BC  & 9167 & 2693 & 1404 & 3545 & 2498 \\
W  & 3216 & 1381 & 2189 & 1916 & 1350 \\
CH  & 3703 & 1493 & 1597 & 2222 & 1859 \\
CL  & 1264 & 771  & 706  & 955  & 1386 \\
\midrule
\textbf{Total}   & 23865 & 8413 & 7963  & 12264  & 11718 \\
\bottomrule
\end{tabular}
}
\label{tab:requests_per_project}
\end{table}

Our results show that across all code adequacy metrics, method-level prompting is substantially more effective than class-level prompting. Although both configurations are provided with the same source of information and the instruction to eventually target identical code, the difference in granularity was significant. We noticed that one contributing factor is the number of generated tests. Under class-level prompting, Plain-LLM generates 3,232 test cases, of which 50.53\% pass successfully. In contrast, method-level prompting produces nearly four times as many tests, generating 12,158 test cases with a slightly higher passing rate of 53.40\%.

\begin{tcolorbox}[colback=gray!10, colframe=gray!50,boxrule=0.4pt,arc=1pt,left=6pt,right=6pt,top=4pt,bottom=4pt,fontupper=\small,before skip=8pt,after skip=8pt,enhanced]
\textbf{Finding 3:} Granularity plays a critical role in both the effectiveness and efficiency of LLM-based test generation. Method-level testing enables the model to explore methods more thoroughly by generating approximately four times as many tests, resulting in higher code coverage.
\end{tcolorbox}

As expected, class-level testing requires significantly fewer LLM queries. Across the dataset, class-level testing generated a total of 1,510 tests, which is approximately 7.76 times fewer than the number generated at the method level.

\begin{table*}[t!]
\centering
\caption{Comparison of Class-level and Method-level testing}
\resizebox{\textwidth}{!}{
\begin{tabular}{lccc|ccc|ccc}
\toprule
\textbf{Project} & 
\multicolumn{3}{c}{\textbf{Line Coverage}} &
\multicolumn{3}{c}{\textbf{Branch Coverage}} &
\multicolumn{3}{c}{\textbf{Mutation Score}} \\
\cmidrule(r){2-4} \cmidrule(r){5-7} \cmidrule(r){8-10}
& \textbf{Class Level} & \textbf{Method Level} & \textbf{Combined}
& \textbf{Class Level} & \textbf{Method Level} & \textbf{Combined}
& \textbf{Class Level} & \textbf{Method Level} & \textbf{Combined} \\
\midrule
BPG & 36.47\% & \underline{45.13\%} & \textbf{49.07\%}
    & 22.88\% & \underline{32.60\%} & \textbf{36.05\%}
    & 29.33\% & \underline{41.21\%} & \textbf{42.59\%} \\
ER  & 35.58\% & \underline{45.55\%} & \textbf{49.89\%}
    & 23.36\% & \underline{32.10\%} & \textbf{35.71\%}
    & 23.03\% & \underline{31.53\%} & \textbf{35.24\%} \\
BC  & 15.21\% & \underline{49.91\%} & \textbf{51.06\%}
    & 36.20\% & \underline{49.77\%} & \textbf{53.39\%}
    & 7.85\%  & \underline{15.93\%} & \textbf{16.39\%} \\
W   & \underline{37.14\%} & 36.96\% & \textbf{44.72\%}
    & 24.90\% & \underline{27.49\%} & \textbf{31.87\%}
    & 30.48\% & \underline{32.81\%} & \textbf{38.88\%} \\
CM  & 41.62\% & \underline{65.39\%} & \textbf{69.08\%}
    & 25.56\% & \underline{48.89\%} & \textbf{55.37\%}
    & 28.50\% & \underline{53.24\%} & \textbf{57.90\%} \\
CH  & 31.24\% & \underline{61.86\%} & \textbf{63.67\%}
    & 8.57\%  & \underline{39.10\%} & \textbf{40.22\%}
    & 14.11\% & \underline{30.00\%} & \textbf{31.71\%} \\
\midrule
\textbf{Total} & 33.53\% & \underline{49.95\%} & \textbf{53.67\%}
              & 21.05\% & \underline{35.33\%} & 3\textbf{8.74\%}      & 22.08\% & \underline{33.82\%} & \textbf{36.55\%} \\
\bottomrule
\end{tabular}
}
\label{tab:class_method_combined}
\end{table*}

\section{Discussion}

\subsection{Generalization on multiple LLMs}

To assess whether our findings generalize beyond the primary model under evaluation, we replicated our experiments using two additional large language models: DeepSeek V3 and LLaMA 3.3 (70B). These models were selected because they belong to different model families and represent state-of-the-art capabilities in large-scale code generation.

It is important to note, however, that the two models differ in their announced training data cutoff dates. LLaMA 3.3 reports a cutoff of December 2023, whereas DeepSeek V3 reports a substantially later cutoff of December 2024. As a result, DeepSeek V3 may have been exposed to the full set of projects used in our evaluation.

Table \ref{tab:rates_in_all_llms} summarizes the compilation and passing rates across the entire dataset for all evaluated models. Consistent with our earlier observations for GPT-4o-mini, method-level prompting yields substantially more generated tests than class-level prompting. For DeepSeek and LLaMA, we observe a modest improvement in passing rates; however, this comes at the cost of reduced compilation rates, indicating a trade-off between test correctness and syntactic validity in these models’ outputs.

\begin{table}[b]
\centering
\caption{Compilation and Passing Rates between different LLMs}
\scriptsize
\resizebox{\columnwidth}{!}{
\begin{tabular}{lccccc}
\toprule
 & \multicolumn{2}{c}{\textbf{Compilation Rate}} 
 & \multicolumn{2}{c}{\textbf{Passing Rate}} \\
\cmidrule(lr){2-3} \cmidrule(lr){4-5}
\textbf{Name} 
& \textbf{Plain (Class)} 
& \textbf{Plain (Method)} 
& \textbf{Plain (Class)} 
& \textbf{Plain (Method)} \\
\midrule
GPT 4o-mini     
& 88.64\% (2865/3232)  
& 84.34\% (10254/12158)
& 50.53\% (1633/3232)  
& 53.40\% (6492/12158) \\
Deepseek V3     
& 82.47\% (3800/4608)  
& 83.02\% (16539/19922)
& 61.87\% (2851/4608)  
& 65.57\% (13062/19922) \\
Llama 3.3 (70B) 
& 65.56\% (2640/4027)  
& 79.14\% (10477/13239)
& 55.55\% (2237/4027)  
& 57.30\% (7586/13239) \\
\bottomrule
\end{tabular}
}
\label{tab:rates_in_all_llms}
\end{table}

Table \ref{tab:effectiveness_all_models} presents the effectiveness of each approach across all LLMs used. The trends observed previously are consistent across all models, supporting the generalizability of our findings. In particular, for DeepSeek V3, where the entire dataset may have been included in its training corpus, the Plain-LLM approach substantially outperforms all others. Although this cannot be confirmed definitively, the evidence suggests that, over time, the effectiveness of prior approaches may diminish.

\begin{figure}[ht]
    \centering
    \includegraphics[width=\linewidth]{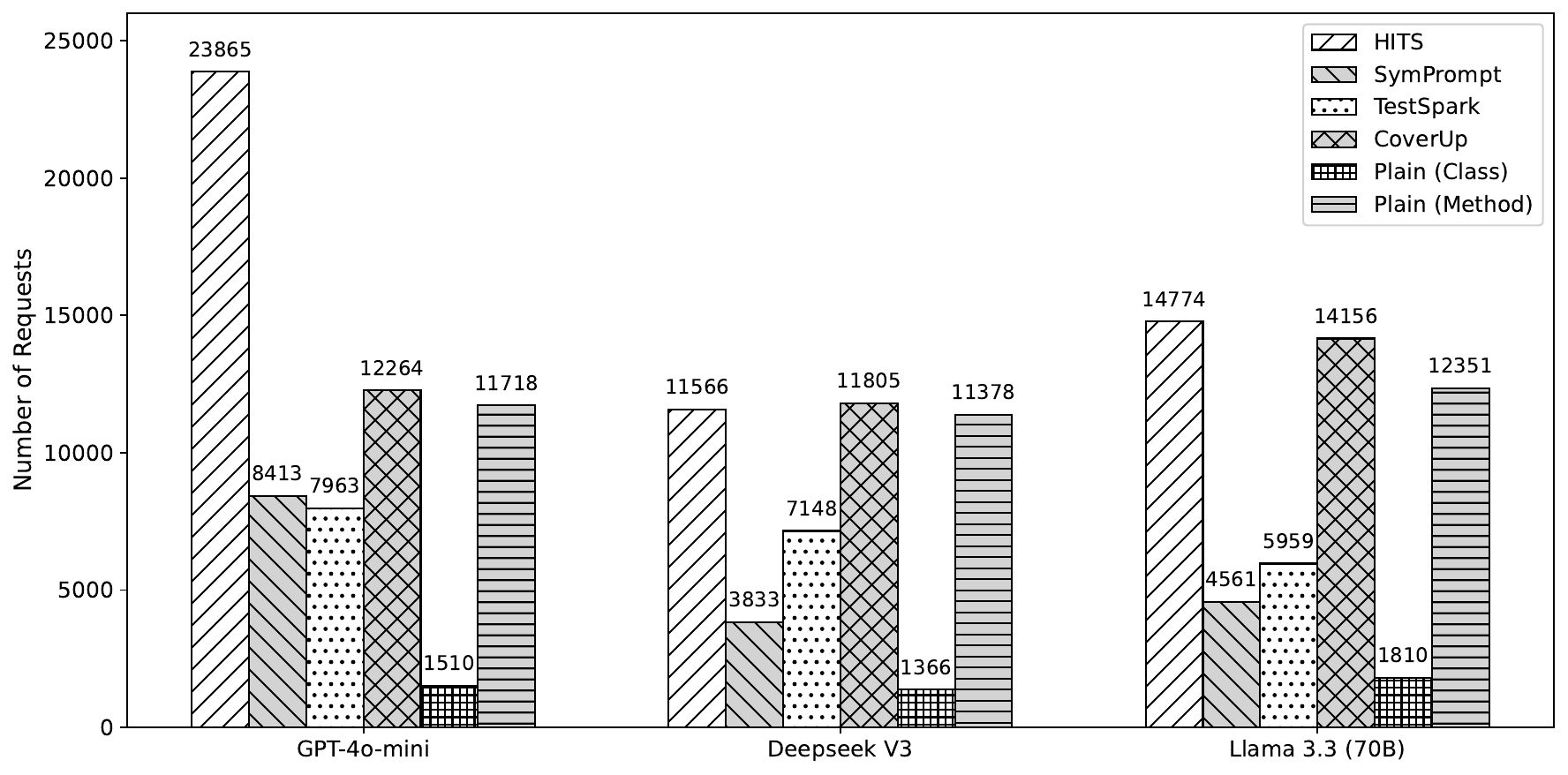}
    \caption{Number of API requests between different approaches}
    \label{fig:requests_per_llm_all_techniques}
\end{figure}

Regarding efficiency, Figure~\ref{fig:requests_per_llm_all_techniques} shows the number of requests of each approach to each LLM, and overall the trend looks similar. 
However, in the case of SymPromt, it clearly generated fewer tests with Llama 3.3 70B and DeepSeek-V3, which explains its degradation in terms of effectiveness. 

We can also observe that the number of requests done by TestSpark, CoverUP, and the Plain-LLM (method and class granularity), remain consistent throughout the different LLMs. HITS, however, performed many more requests to the LLM when gpt-4o-mini is used, which explains why its performance is higher under this specific LLM.

\begin{table*}[t!]
\centering
\caption{Code coverage between different Approaches across all LLMs used}
\footnotesize
\resizebox{\textwidth}{!}{
\begin{tabular}{llcccccc|cccccc|cccccc}
\toprule
 & & \multicolumn{6}{c}{\textbf{Line Coverage}} 
 & \multicolumn{6}{c}{\textbf{Branch Coverage}} 
 & \multicolumn{6}{c}{\textbf{Mutation Score}} \\
\cmidrule(r){3-8} \cmidrule(r){9-14} \cmidrule(r){15-20}
\textbf{Model} & \textbf{Value} 
& \textbf{HITS} & \textbf{SYMP} & \textbf{TSP} & \textbf{CUP} & \textbf{Plain (C)} & \textbf{Plain (M)}
& \textbf{HITS} & \textbf{SYMP} & \textbf{TSP} & \textbf{CUP} & \textbf{Plain (C)} & \textbf{Plain (M)}
& \textbf{HITS} & \textbf{SYMP} & \textbf{TSP} & \textbf{CUP} & \textbf{Plain (C)} & \textbf{Plain (M)} \\
\midrule
\multirow{1}{*}{GPT-4o-mini} & Total   & 42.43\% & 40.95\% & 23.65\% & 38.85\% & 33.53\% & \textbf{49.95\%} & 29.49\% & 27.19\% & 16.48\% & 25.11\% & 21.05\% & \textbf{35.35\%} & 27.97\% & 28.40\% & 13.88\% & 26.26\% & 22.07\% & \textbf{33.81\%} \\
                              
\cmidrule(lr){1-20}
\multirow{1}{*}{Deepseek V3} & Total   & 48.48\% & 33.74\% & 25.24\% & 42.10\% & 38.91\% & \textbf{59.82\%} & 36.40\% & 22.09\% & 19.51\% & 32.98\% & 23.32\% & \textbf{49.95\%} & 30.92\% & 16.05\% & 16.60\% & 28.17\% & 23.88\% & \textbf{46.05\%} \\
\cmidrule(lr){1-20}
\multirow{1}{*}{Llama 3.3 (70B)} & Total   & 41.84\% & 28.56\% & 25.00\% & 36.61\% & 43.73\% & \textbf{52.34\%} & 31.05\% & 16.38\% & 18.54\% & 25.43\% & 28.60\% & \textbf{37.84\%} & 26.63\% & 13.37\% & 14.19\% & 20.51\% & 28.68\% & \textbf{36.53\%} \\
                                   
\bottomrule
\end{tabular}
}
\label{tab:effectiveness_all_models}
\end{table*}

\begin{tcolorbox}[colback=gray!10, colframe=gray!50,boxrule=0.4pt,arc=1pt,left=6pt,right=6pt,top=4pt,bottom=4pt,fontupper=\small,before skip=8pt,after skip=8pt,enhanced]
\textbf{Finding 4:} Our findings generalize across multiple LLMs from different model families. Experiments with LLaMA 3.3 (70B) yield results consistent with those observed previously, and evaluations using DeepSeek V3 reveal an even larger performance gap between Plain-LLM and prior approaches.
\end{tcolorbox}

\subsection{Method and Class Granularity Complementarity}

Although the generated class-level suites may look less effective than the method-level ones, we observed that these are indeed complementary. We merged the passing tests generated at the class-level and method-level from the previous experiment into a single test suite. 

As shown in Table \ref{tab:class_method_combined}, this combined suite consistently achieves higher coverage across all evaluated code coverage metrics. This observation indicates that the two approaches are complementary. While method-level prompting tends to produce more fine-grained and in-depth tests, class-level prompting enables the model to exercise code outside individual methods, thereby improving overall coverage. 

\begin{table}[t]
\centering
\caption{Comparison of Combined and Hybrid results in terms of coverage}
\scriptsize
\resizebox{\columnwidth}{!}{
\begin{tabular}{lcc|cc|cc}
\toprule
\textbf{Project} & 
\multicolumn{2}{c}{\textbf{Line Coverage}} &
\multicolumn{2}{c}{\textbf{Branch Coverage}} &
\multicolumn{2}{c}{\textbf{Mutation Score}} \\
\cmidrule(r){2-3} \cmidrule(r){4-5} \cmidrule(r){6-7}
& \textbf{Combined} & \textbf{Hybrid} 
& \textbf{Combined} & \textbf{Hybrid} 
& \textbf{Combined} & \textbf{Hybrid} \\
\midrule
BPG & 49.07\% & 49.47\% 
    & 36.05\% & 38.56\% 
    & 42.59\% & 44.27\% \\
ER  & 49.89\% & 44.30\% 
    & 35.71\% & 33.03\% 
    & 35.24\% & 31.39\% \\
BC  & 51.06\% & 47.86\% 
    & 53.39\% & 54.30\% 
    & 16.39\% & 17.10\% \\
W   & 44.72\% & 43.98\% 
    & 31.87\% & 31.87\% 
    & 38.88\% & 37.17\% \\
CM  & 69.08\% & 71.47\% 
    & 55.37\% & 62.22\% 
    & 57.90\% & 60.36\% \\
CH  & 63.67\% & 63.98\% 
    & 40.22\% & 40.14\% 
    & 31.71\% & 35.70\% \\
\midrule
\textbf{Total} & 53.67\% & 52.30\% 
                & 38.74\% & 38.84\% 
                & 36.55\% & 36.76\% \\
\bottomrule
\end{tabular}
}
\label{tab:combined_vs_hybrid}
\end{table}

\begin{tcolorbox}[colback=gray!10, colframe=gray!50,boxrule=0.4pt,arc=1pt,left=6pt,right=6pt,top=4pt,bottom=4pt,fontupper=\small,before skip=8pt,after skip=8pt,enhanced]
\textbf{Finding 5:} Class-level and method-level workflows are complementary. Combining the benefits of the two, one can achieve higher coverage. 
\end{tcolorbox}

Nonetheless, combining class-level and method-level test suites requires a substantially higher cost in terms of generated tests and execution time, and increases the likelihood of duplicate tests. To address these limitations, we propose a \textit{hybrid approach} that integrates the strengths of both workflows while improving efficiency. The \textit{hybrid approach} initially applies class-level test generation and subsequently invokes method-level test generation only for methods that are not fully covered at the branch level. In other words, method-level testing is selectively applied to insufficiently covered methods.

Table \ref{tab:combined_vs_hybrid} compares the results of the combined and hybrid approaches showing that the observed differences are minimal. Statistical analysis using the Mann–Whitney U test~\cite{mann1947test} confirms that these differences are not statistically significant (p = 0.92 for branch coverage and p = 0.75 for line coverage). In other words, these results indicate that there is no difference in terms of effectiveness between combined and hybrid approach. 

When it comes to efficiency though, our findings show that the hybrid approach is more optimized. In table \ref{tab:combined_vs_hybrid_requests_tests} we observe that hybrid approach requires fewer queries to the LLM and used approximately half the number of tests compared to the combined approach. Furthermore, hybrid approach not only outperformed the combined test suite, but it was even more efficient than using only method-level testing. 

\begin{figure}[ht]
    \centering
    \includegraphics[width=\linewidth]{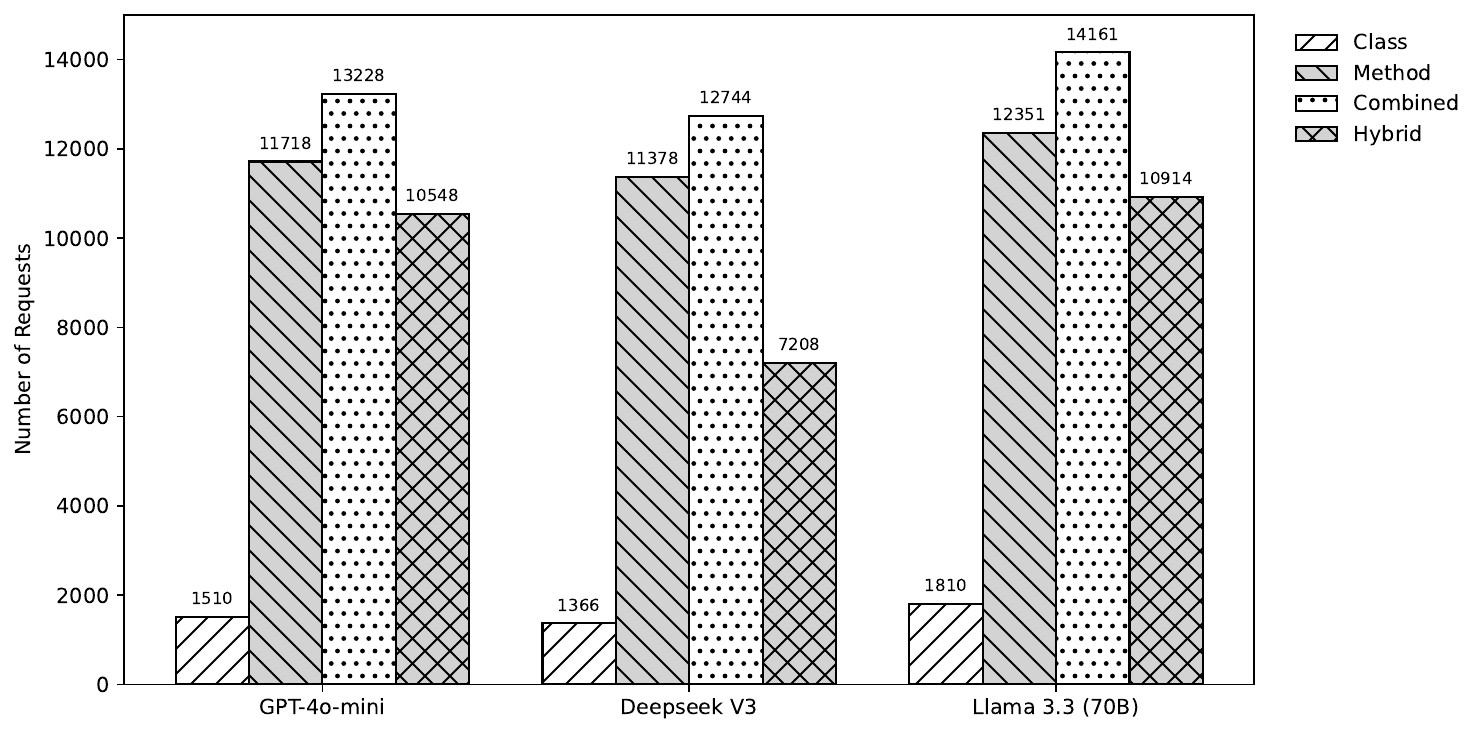}
    \caption{Number of API requests for LLM-Plain across multiple LLMs}
    \label{fig:requests_per_llm_plain}
\end{figure}

Figure \ref{fig:requests_per_llm_plain} depicts the number of queries issued by LLM-Plain to each respective model. Consistent with our previous observations, class-level prompting requires the fewest queries, whereas method-level and combined-level prompting necessitate substantially more. Notably, the Hybrid approach provides a favorable trade-off throughout the three LLMs, achieving improved coverage while reducing the total number of API requests. 

\begin{tcolorbox}[colback=gray!10, colframe=gray!50,boxrule=0.4pt,arc=1pt,left=6pt,right=6pt,top=4pt,bottom=4pt,fontupper=\small,before skip=8pt,after skip=8pt,enhanced]
\textbf{Finding 6:} Hybrid approach achieves the same effectiveness as the combined class-level and method-level test suites, while exhibiting greater efficiency by requiring fewer LLM queries and generating fewer tests.
\end{tcolorbox}

\begin{table}[h]
\centering
\caption{Comparison of Combined and Hybrid in terms of \# Requests and \# Tests (passing)}
\scriptsize
\begin{tabular}{lrr|rr}
\toprule
\textbf{Project} & 
\multicolumn{2}{c}{\# Requests} &
\multicolumn{2}{c}{\# Tests} \\
\cmidrule(r){2-3} \cmidrule(r){4-5}
& \textbf{Combined} & \textbf{Hybrid} 
& \textbf{Combined} & \textbf{Hybrid} \\
\midrule
BPG & 2524 & \textbf{2021} & 2877 & \textbf{1309} \\
ER  & 2525 & \textbf{2310} & 1149 & \textbf{690} \\
BC  & 2749 & \textbf{2682} & 405  & \textbf{336} \\
W   & 1703 & \textbf{1309} & 827  & \textbf{464} \\
CM  & 2189 & \textbf{1437} & 980  & \textbf{693} \\
CH  & 1538 & \textbf{789}  & 1887 & \textbf{834} \\
\midrule
\textbf{Total} & 13228 & \textbf{10548} & 8125 & \textbf{4326} \\
\bottomrule
\end{tabular}
\label{tab:combined_vs_hybrid_requests_tests}
\end{table}

\subsection{High number of non-compiling and failing test cases}

Although all evaluated approaches aim to achieve full coverage, our results show that this goal remains largely out of reach. A central challenge across all approaches is the high number of non-compiling and failing test cases. As reported in Table \ref{tab:rates_in_all_llms}, a substantial portion of the generated tests either fail to compile or fail at runtime due to incorrect oracles.

In addition, based on our manual inspection, several failing cases cannot be attributed solely to incorrect oracles. In multiple instances, the failures stem from improperly mocked external dependencies. However, distinguishing between failures caused by incorrect assertions and those caused by an incorrect test prefix is difficult. To better understand these shortcomings, we conducted a detailed analysis of the generated test cases in order to identify common sources of failure.

\subsubsection{Syntactically invalid test cases}

In the implementation of ChatUniTest, which we use to execute the four state-of-the-art tools, non-compiling tests are filtered out by design if the execution is successful. For certain approaches, such as CoverUp, this behavior is unavoidable, as CoverUp requires a fully passing test suite to compute coverage for the class under test. In contrast, when running LLM-Plain, we are able to directly measure the number of non-compiling tests. This analysis reveals that a key limitation of several approaches lies in the generation of syntactically invalid test cases. 

Table \ref{tab:rates_in_all_llms} reports the compilation and passing rates of LLM-Plain at both the class-level and method-level across all evaluated LLMs. On average, 78.89\% of the generated tests successfully compile at the class level, compared to 82.17\% at the method level. With respect to test execution, approximately half of the generated tests pass: on average, 55.98\% at the class level and 59.42\% at the method level. These results indicate that, despite LLMs’ ability to generate test code, only about 40–50\% of the generated tests are ultimately usable. 

\begin{tcolorbox}[colback=gray!10, colframe=gray!50,boxrule=0.4pt,arc=1pt,left=6pt,right=6pt,top=4pt,bottom=4pt,fontupper=\small,before skip=8pt,after skip=8pt,enhanced]
\textbf{Finding 6:} Plain-LLM demonstrates that one of the key reasons limiting the effectiveness of LLM-based test generation approaches is the frequent generation of syntactically and semantically invalid test cases. In our evaluation, only about half of the tests generated by LLM-Plain successfully passed.
\end{tcolorbox}

Although not all corresponding studies explicitly report passing rates, the results that are available align with our findings. For instance, the passing rate reported for HITS closely matches the passing rate observed in our evaluation.

\subsubsection{Code generation beyond unit tests}

One of the earliest observations from our analysis was that a common cause of non-compiling or failing tests was the generation of additional classes or interfaces by the LLM. In several cases, instead of correcting the unit tests themselves, the LLM attempted to resolve compilation or runtime errors by introducing new classes or interfaces intended to override existing functionality. 
Hallucinations are a known prevalent issue on LLMs~\cite{LLMHallucinationsSurvey}.

\begin{figure}[ht]
    \centering
    \includegraphics[width=\linewidth]{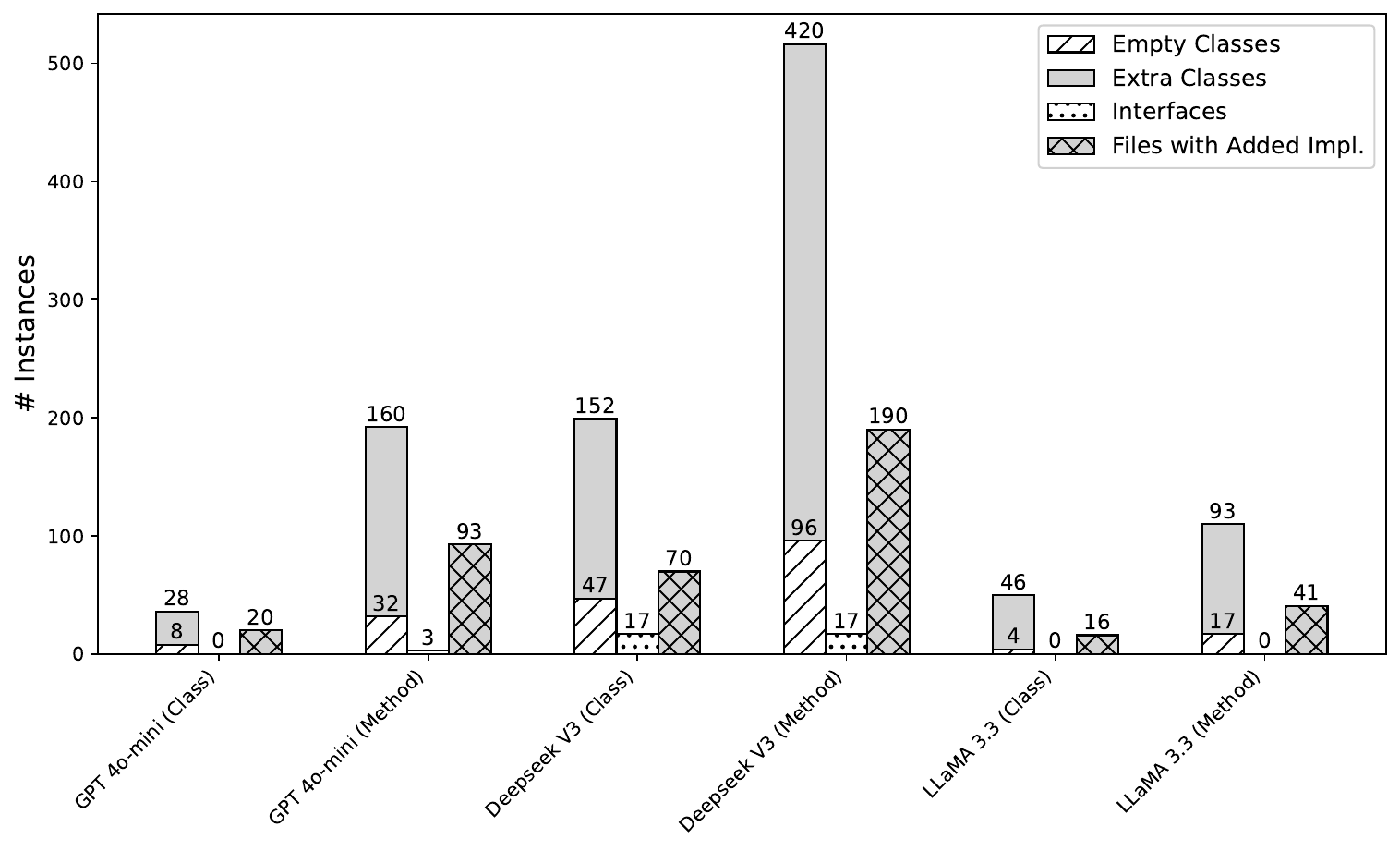}
    \caption{Additional content generated by LLMs instead of fixing unit tests}
    \label{fig:additional_content}
\end{figure}

To better understand the prevalence of this behavior, we conducted a systematic analysis of the generated files and quantified the number of occurrences. Figure \ref{fig:additional_content} illustrates both the number of instances in which the LLM generated additional classes or interfaces and the number of test files in which the LLM attempted to replace parts of the existing implementation. Such limitation shows that we need to find better test-repair mechanisms. 

Nonetheless, we observed another noteworthy behavior: the generation of \textit{empty class}. Listing \ref{lst:additional_class_by_deepseek} presents an example of such an \textit{empty class}. As shown, the generated class contains no functionality and appears to be introduced by the LLM merely as a placeholder.

\begin{lstlisting}[float, floatplacement=t,
    caption={Example of additional yet empty class generated by DeepSeek},
    label={lst:additional_class_by_deepseek},
    captionpos=b]
/** Helper classes for testing */
private static class TestClass {
    // Simple test class
}
\end{lstlisting}

Figure \ref{fig:additional_content} further quantifies this phenomenon, indicating that empty classes account for approximately 10–30\% of all additionally generated classes.

\begin{tcolorbox}[colback=gray!10, colframe=gray!50,boxrule=0.4pt,arc=1pt,left=6pt,right=6pt,top=4pt,bottom=4pt,fontupper=\small,before skip=8pt,after skip=8pt,enhanced]
\textbf{Finding 7:} LLMs tend to generate additional code beyond the intended unit tests. Across all evaluated models, we observed the generation of extra interfaces and auxiliary classes, as well as empty classes that appear to serve as placeholders.
\end{tcolorbox}

\begin{figure}[ht]
    \centering
    \includegraphics[width=\linewidth]{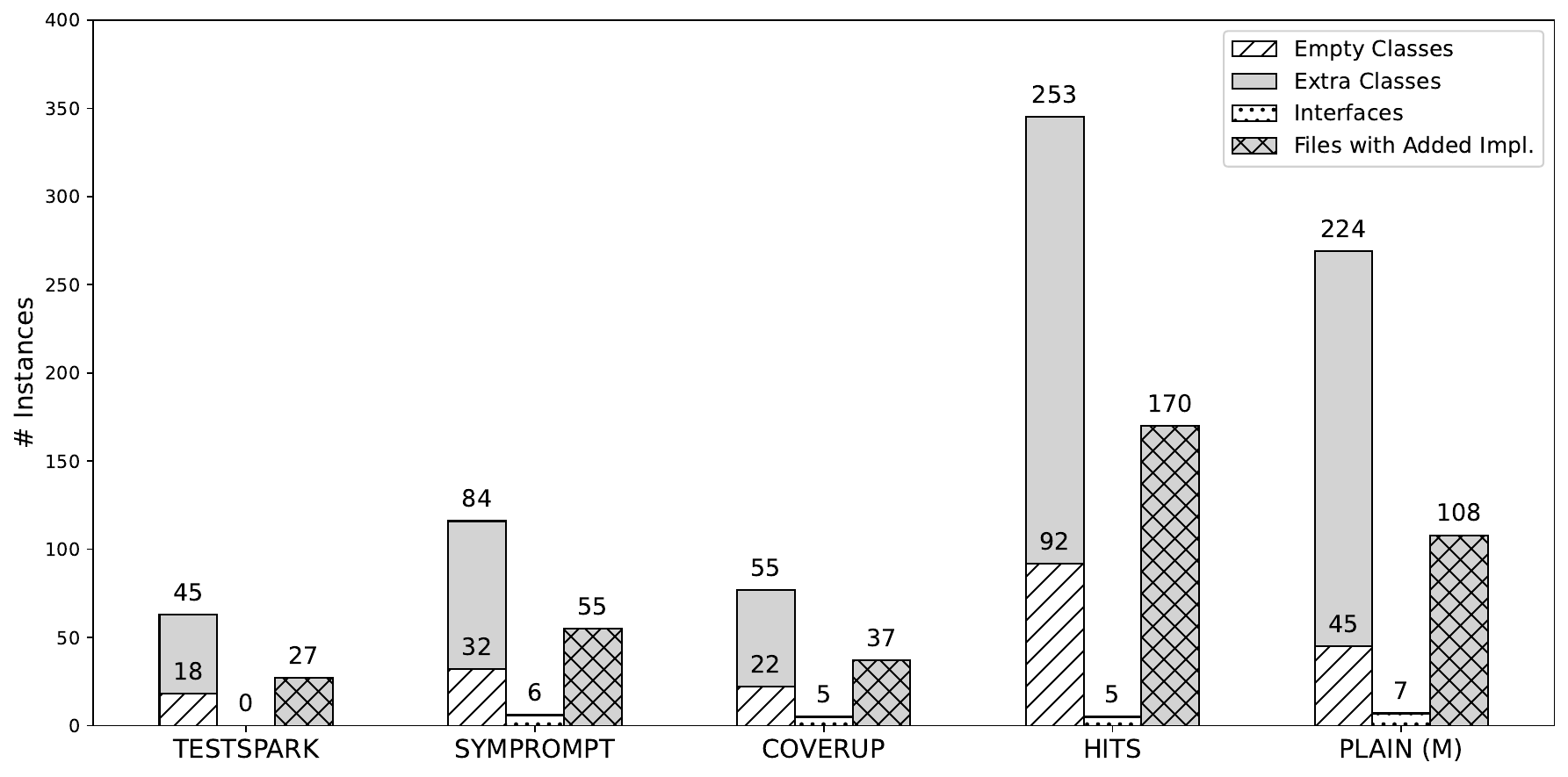}
    \caption{Average additional content generated on whole dataset}
    \label{fig:test_file_analysis_per_algo}
\end{figure}

Curious to see whether this phenomenon occurs in other approaches as well, we calculated all additional generated interfaces and classes in all approaches across the entire dataset. Our analysis confirms that all evaluated approaches produce extra content. Specifically, HITS, SymPrompt, CoverUp, TestSpark, and our implementation of LLM-Plain generated additional interfaces or classes in 6.01\%, 5.14\%, 3.89\%, 6.24\%, and 3.98\% of all generated files, respectively. Listing \ref{lst:additional_helper_hits} shows an additional class generated by HITS, which overrides existing content to assist on its generated tests. Figure \ref{fig:test_file_analysis_per_algo} provides a detailed breakdown of the number of additional generated instances for each approach across the full dataset.

\begin{lstlisting}[float, floatplacement=t, caption={Example of additional generated by HITS},
    label={lst:additional_helper_hits},
    captionpos=b]
    // Test class for JSON deserialization
    static class User {
        [...]
        @Override
        public boolean equals(Object o) {
            if (this == o)
                return true;
            if (o == null || getClass() != o.getClass())
                return false;
            User user = (User) o;
            return age == user.age && (name != null ? name.equals(user.name) : user.name == null);
        }

        @Override
        public int hashCode() {
            int result = name != null ? name.hashCode() : 0;
            result = 31 * result + age;
            return result;
        }
    }
\end{lstlisting}

\begin{tcolorbox}[colback=gray!10, colframe=gray!50,boxrule=0.4pt,arc=1pt,left=6pt,right=6pt,top=4pt,bottom=4pt,fontupper=\small,before skip=8pt,after skip=8pt,enhanced]
\textbf{Finding 8:} All evaluated LLM-based approaches generated additional content beyond unit tests, which either overrides existing functionality, implements supplementary test classes, or serves as placeholders for developers to insert necessary code.
\end{tcolorbox}

\section{Open challenges}


\subsection{Fixing syntactically invalid tests}
Our findings indicate that a substantial number of generated tests are discarded because they either fail to compile or do not pass, potentially due to incorrectly generated oracles. Averaging across all evaluated models, we found that at the class level, only 78.89\% of tests successfully compiled, while on method level about 82,17\%. In other words, an average of 854 tests per model were discarded at the class level, and 2,683 tests per model at the method level.

Although all approaches in our evaluation discard non-compiling tests, it is possible that an approach incorporating a test-repair mechanism could achieve improved results by correcting otherwise discarded tests.

Similarly, the prevalence of non-passing tests cannot be overlooked, as on average approximately half of the generated tests fail, thereby reducing the overall effectiveness of the generated test suite. Perhaps the primary reason for the failure of otherwise compiling tests appears to be incorrectly generated oracles. Using LLMs for test oracle generation is a research challenge in its own, with distinct advantages and limitations~\cite{molina2025test}~\cite{konstantinou2024llms}. Nonetheless, some approaches have already combined traditional test generation techniques with LLM-based oracle generation~\cite{dinella2022toga}~\cite{hossain2024togll}. A dedicated line of research that integrates an oracle generation mechanism into LLM-based test generation tools could substantially improve the quality and effectiveness of the generated tests.

\subsection{Proper guidance: Avoid creativity and filling helpers}

Based on our case study, we observed that a substantial number of classes could not be effectively tested using straightforward unit tests alone. Some classes required auxiliary helper classes to enable meaningful test execution. Although the LLMs occasionally attempted to generate such helper classes, these attempts were largely unsuccessful, leading us to exclude the generated helpers from the evaluation.

Moreover, beyond failing to produce functional helper classes, we found that approximately 10–30\% of the generated classes were placeholders lacking concrete implementations and seemingly intended for manual completion by a developer. In other cases, the LLMs generated classes that overrode existing functionality rather than supporting it.

While overriding production functionality is generally undesirable, these observations suggest that automated test generation could benefit from the ability to synthesize accurate and well-scoped helper components. To the best of our knowledge, this aspect of LLM-based test generation has not yet been explored and represents a promising direction for future work.

\subsection{Hybrid approach: Leveraging granularity}

Our experiments show that granularity has a significant impact on effectiveness, as method-level prompting makes LLMs to generate more tests and cover a greater number of execution paths. At the same time, the two approaches (class-level and method-level) do not replace each other, but rather complement one another. However, while combining class-level and method-level approaches is effective, it is relatively cost-inefficient, requiring a high number of LLM queries and increasing the likelihood of generating duplicate tests. 

In our case study, we observed that the Hybrid approach achieves effectiveness comparable to the combined approach while substantially reducing LLM query cost, despite our current implementation not being fully optimized. Moreover, whereas prior approaches in our evaluation primarily rely on method-level testing, our findings indicate that the Hybrid approach can be even more efficient than method-level testing alone. On average across all evaluated models, the Hybrid approach required 2,259 fewer LLM queries than method-level testing, which is the strategy employed by current approaches. These results highlight the need for further research on efficiently combining multiple levels of granularity to maximize effectiveness and efficiency.

\section{Threads to Validity}

\textbf{External validity: } A potential threat to the validity of this study concerns the tools used for comparison. To mitigate this risk, we relied on ChatUniTest, a widely adopted plugin that implements all the algorithms included in our evaluation. This tool has been employed in previous studies, including evaluations of LLM-based test generation approaches. In addition, we evaluated multiple LLMs from different model families to ensure that our findings are consistent and generalizable. 

\textbf{Internal validity: } The performance of our implementation depends on the underlying LLMs. Accordingly, a potential threat to validity arises from variations in model capability as well as from the nondeterministic nature of LLM outputs. To mitigate these threats, we evaluated multiple models from different families, reducing the risk that our findings are specific to a single model. Additionally, to limit nondeterminism, we configured all models with a low temperature setting (0.1), aiming to make the generation process as deterministic as possible.

\textbf{Construct validity: } A potential threat to the validity of this study is data leakage, that is, whether the LLM has already been exposed to our evaluation set during training. To mitigate this threat, we selected repositories previously used in studies on LLM-based test generation~\cite{chatunitest/10.1145/3663529.3663801}~\cite{WangL0J24}, as well as the GitBug-Java dataset~\cite{SilvaSM24}. Additionally, we manually inspected the creation date of each repository included in our experimental setup and labeled them accordingly to distinguish between seen and unseen data.

\section{Conclusion}

In this study, we investigated the performance of existing LLM-based test generation approaches when evaluated on newer LLM versions than those used in their original assessments. We also implemented a simple \textit{Plain-LLM} baseline that relies exclusively on the capabilities of the LLM, employing minimal output post-processing and no error handling or sophisticated engineering mechanisms.

Our findings indicate that simple prompting via Plain-LLM can outperform the four state-of-the-art tools included in our evaluation. Plain-LLM achieved higher total and average coverage than all prior approaches across most metrics, with the only exception being that HITS slightly outperformed Plain-LLM on average branch coverage. Regarding efficiency, Plain-LLM demonstrated comparable performance.

We further examined the impact of prompting granularity on LLM-based test generation. Our findings show that method-level prompting leads LLMs to generate a larger number of tests and to exercise methods more thoroughly than class-level prompting. To explore whether these two strategies can be combined, we evaluated their joint use and found them to be complementary. Building on this observation, we demonstrated that a hybrid approach that leverages both class-level and method-level workflows can improve the cost-effectiveness of test generation, outperforming method-level prompting alone as well as the naïve combination of their respective test suites.

Finally, we demonstrated that these results generalize across multiple LLMs of different families. In addition, we observed that LLM-based test generation techniques suffer from a large number of syntactical invalid tests, as under class level, only 78.89\% of tests successfully compiled, while on method level about 82,17\%.

\section{Data availability}

We provide a complete and executable implementation of LLM-Plain at the following link: 

\url{https://github.com/michaelkonstantinou/llm-plain}

The repositories included in our dataset, as well as the state-of-the-art tools employed, are publicly available.

\section*{Acknowledgment}

This work is supported by the Luxembourg National Research Fund (FNR), Grant number C22/IS/17426831/MeMoRIA; the FNR PEARL program Grant agreement 16544475; and the RDI Law project ``Innovations for 21st Century Assessment Authoring,'' financed by the Luxembourg Ministry of the Economy.

\bibliographystyle{IEEEtran}
\bibliography{main}

\end{document}